\documentclass[aps, superscriptaddress, preprintnumbers, prd, twocolumn]{revtex4}

\usepackage{amsmath}	
\usepackage{amsthm}	
\usepackage{amssymb}	
\usepackage{datetime}
\usepackage{graphicx}
\usepackage{color}
\usepackage{verbatim}
\usepackage{bm}

\usepackage{hyperref}

\definecolor{grey}{rgb}{0.4,0.4,0.4}
\definecolor{dullmagenta}{rgb}{0.4,0,0.4}
\definecolor{darkblue}{rgb}{0,0,0.4}
\definecolor{midblue}{rgb}{0,0,0.5}
\definecolor{midred}{rgb}{0.5,0,0}
\definecolor{orange}{rgb}{1,0.5,0}
\definecolor{lightbrown}{rgb}{0.75,0.5,0.25}
\definecolor{tan}{cmyk}{0.14,0.42,0.56,0}
\definecolor{djunglegreen}{cmyk}{0.99,0,0.52,0}
\definecolor{lightgreen}{rgb}{0,1,0}
\definecolor{olivegreen}{cmyk}{0.64,0,0.95,0.40}
\definecolor{midgreen}{rgb}{0.0,0.675,0.0}
\definecolor{darkgreen}{rgb}{0,0.5,0}

\newcommand{\vs}{\vspace}

\renewcommand{\.}{\hspace{0.5mm}}

\newcommand{\ra}{\ensuremath{\rightarrow}}

\newcommand{\Nbb}{\ensuremath{\mathbb{N}}}

\renewcommand{\d}{\ensuremath{\mathrm{d}}}

\newcommand{\ie}{ie.}

\newcommand{\rhs}{r.h.s.}

\newcommand{\cf}{cf.}

\let\baraccent=\= 
\renewcommand{\=}[1]{\stackrel{#1}{=}} 

\theoremstyle{definition}

\theoremstyle{remark}

\settimeformat{ampmtime}

\usepackage{float}

\begin{document}

\title{Constraints on Primordial Black Holes with Extended Mass Functions}

\author{Florian K{\"u}hnel}
\email{florian.kuhnel@fysik.su.se}
\affiliation{The Oskar Klein Centre for Cosmoparticle Physics,
	Department of Physics,
	Stockholm University,
	AlbaNova University Center,
	Roslagstullsbacken 21,
	SE--106\.91 Stockholm,
	Sweden}

\affiliation{Department of Physics,
	School of Engineering Sciences,
	KTH Royal Institute of Technology,
	AlbaNova University Center,
	Roslagstullsbacken 21,
	SE--106\.91 Stockholm,
	Sweden}

\author{Katherine Freese}
\email{ktfreese@umich.edu}
\affiliation{Department of Physics,
	University of Michigan,
	Ann Arbor,
	MI 48109,
	USA}

\affiliation{The Oskar Klein Centre for Cosmoparticle Physics,
	Department of Physics,
	Stockholm University,
	AlbaNova University Center,
	Roslagstullsbacken 21,
	SE--106\.91 Stockholm,
	Sweden}
	
\date{\formatdate{\day}{\month}{\year}, \currenttime}

\begin{abstract}
Constraints on primordial black holes in the range $10^{-18}\.M_{\odot}$ to $10^{3}\.M_{\odot}$ are reevaluated for a general class of extended mass functions. Whereas previous work has assumed that PBHs are produced with one single mass, instead there is expected to be a range of masses even in the case of production from a single mechanism; constraints therefore change from previous literature.
Although tightly constrained in the majority of cases, it is shown that, even under conservative assumptions, primordial black holes in the mass range $10^{-10}\.M_{\odot}$ to $10^{-8}\.M_{\odot}$ could still constitute the entirety of the dark matter. This stresses both the importance for a comprehensive reevaluation of all respective constraints that have previously been evaluated only for a monochromatic mass function, and the need to obtain more constraints in the allowed mass range.\\
\end{abstract}

\maketitle

\noindent
{\it Introduction\;---\;}Primordial black holes (PBHs) are black holes produced in the early Universe, and have received considerable attraction since they were proposed more than four decades ago \cite{1967SvA....10..602Z, Carr:1974nx}. In principle, they can span a huge range of mass scales{\,---\,}from as low as the Planck mass to many orders of magnitude above the solar mass. They are unique probes of the amplitude of the density fluctuation on the very small scales of their production.

With the milestone discovery of the LIGO and VIRGO collaboration of black-hole binary mergers \cite{Abbott:2016blz, Abbott:2016nmj}, the interest in the question of whether PBHs could constitute the dark matter (DM) \cite{1975Natur.253..251C} has recently been revived \cite{Jedamzik:1996mr, Niemeyer:1997mt, Jedamzik:2000ap, Frampton:2010sw, Capela:2012jz, Griest:2013aaa, Belotsky:2014kca, Young:2015kda, Frampton:2015xza, Bird:2016dcv, Kawasaki:2016pql, Carr:2016drx, Kashlinsky:2016sdv, Clesse:2016vqa, Green:2016xgy}. Depending on the mass scale(s) involved, the black holes potentially cause large observable effects. For the case where all PBHs have one single mass (monochromatic mass function) somewhere in the range $10^{-18}\.M_{\odot}$ to $10^{3}\.M_{\odot}$, Figure \ref{fig:Constraints} summarises the strongest constraints from previous literature at each value of possible PBH mass. The figure caption lists the physical effects and provides respective references.

Most of the constraints derived in the literature, including those in Fig.~\ref{fig:Constraints}, are subject to the (at best over-simplifying) assumption that PBH formation occurs mono-chromatically, \ie~at one particular mass scale only, despite the fact that PBH mass spectra are generically extended due to the nature of the gravitational collapse leading to their formation \cite{Choptuik:1992jv, Koike:1995jm, Niemeyer:1999ak, Musco:2004ak, Musco:2008hv, Musco:2012au, Kuhnel:2015vtw}. This incorrect assumption can lead to large errors in the prediction of the PBH abundance (\cf~Ref.~\cite{Kuhnel:2015vtw}). {It should be stressed that, given the current constraints displayed in Fig.~\ref{fig:Constraints}, any monochromatic scenario of $100\%$ {\rm PBH} dark matter is strictly excluded. This statement crucially relies on the validity of the constraints used in this work. However, some of these have been disputed. For instance, those deriving from neutron-star capture of black holes \cite{Capela:2013yf} are challenged due to uncertainties in the amount of dark matter inside globular clusters (\cf~Ref.~\cite{Kusenko:2013saa, Inomata:2017okj}). Furthermore, the bounds from PBH accretion \cite{Ricotti:2007au, Ali-Haimoud:2016mbv, Horowitz:2016lib} rely on highly model-dependent and relatively insecure assumptions (spherically-symmetric Bondi accretion etc.). Also, they depend on the so-called duty-cycle parameter whose exact value varies significantly in the relevant literature. Hence, monochromatic scenarios of $100\%$ of PBH dark matter might still be allowed. In this article, however, our point is to investigate whether{\,---\,}even if these stringent constraints (to which we will refer as 'conservative') are taken at face value{\,---\,}extended mass functions still allow for the possibility that the entirety of the dark matter consists of PBHs.

{In this Letter we reconsider the bounds on primordial black-hole dark matter in the range $10^{-18}\.M_{\odot}$ to $10^{3}\.M_{\odot}$ for a wide class of extended mass functions.} All realistic cases with extended mass spectra require a rederivation of the constraints and an integration of these over the whole mass range; this will be done in the present work.\\[-3mm]

\begin{figure}
	\vs{-3mm}
	\centering
	\includegraphics[scale=1,angle=0]{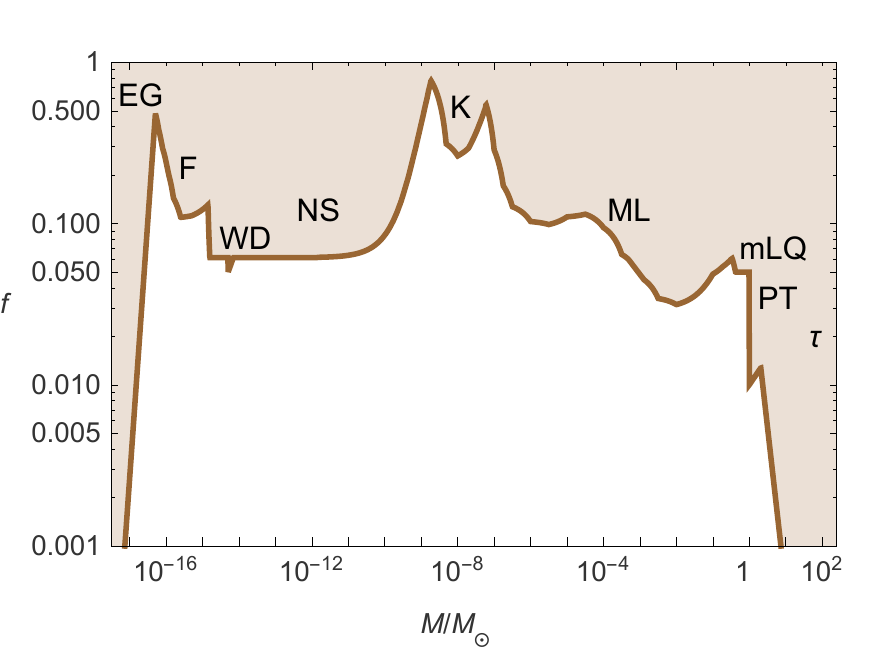}\;
	\vs{-6mm}
	\caption{
		Summary of previous literature for the idealised case in which the entire PBH dark matter 
		consists of PBHs of a single mass $M$ (mono-chromatic mass function):
		Constraint ``curtain'' on the dark-matter fraction $f \equiv \rho_{\rm PBH} / \rho_{\rm DM}$ 
		for a variety of effects associated with PBHs of mass $M$ in units of solar mass $M_{\odot}$. 
		Only strongest constraints are included.
		We show constraints from 
		extragalactic $\gamma$-rays from evaporation (EG) \cite{Carr:2009jm},
		femtolensing of $\gamma$-ray bursts (F) \cite{Barnacka:2012bm}, 
		white-dwarf explosions (WD) \cite{2015PhRvD..92f3007G},
		neutron-star capture (NS) \cite{Capela:2013yf}, 
		Kepler microlensing of stars (K) \cite{Griest:2013aaa}, 
		MACHO/EROS/OGLE microlensing of stars (ML) \cite{Tisserand:2006zx}
		and quasar microlensing (ML) \cite{2009ApJ...706.1451M}, 
		millilensing of quasars (mLQ) \cite{Wilkinson:2001vv}, 
		pulsar timing (PT) (SKA forecast) \cite{Schutz:2016khr},
		and accretion effects on the optical thickness ($\tau$) \cite{Ricotti:2007au}. 
		More details can be found in \cite{Carr:2016drx}.
		Whereas these constraints are only valid for the case of a mono-chromatic mass function, instead
		all realistic cases with extended mass spectra require a rederivation of those constraints 
		and an integration of these over the whole mass range; 
		this will be done in the present work.}
	\label{fig:Constraints}
\end{figure}

\noindent
{\it Primordial Black-Hole Formation\;---\;}There is a plethora of scenarios which lead to the formation of PBHs. All of these require a mechanism to generate large overdensities. In many scenarios, these overdensities are of inflationary origin \cite{Hodges:1990bf, Carr:1993aq, Ivanov:1994pa}. After reentering the horizon, they collapse if they are larger than a (medium-dependent) threshold, where the case of radiation domination is the one most often considered in the literature. Many more possibilities for PBH formation exist, such as those where the source of the inhomogeneities are first-order phase transitions \cite{Jedamzik:1996mr}, bubble collisions \cite{Crawford:1982yz, Hawking:1982ga}, the collapse of cosmic strings \cite{Hogan:1984zb, Hawking:1987bn}, necklaces \cite{Matsuda:2005ez} or domain walls \cite{Berezin:1982ur}.

As mentioned above, in general, one encounters extended PBH mass spectra rather than PBHs of a single mass. The reason is two-fold: On the one hand, the initial spectrum of overdensities is already extended in essentially all of the models mentioned above. On the other hand, even if the initial density spectrum was mono-chromatic, the phenomenon of critical collapse \cite{Niemeyer:1997mt} will inevitably lead to a PBH mass spectrum which is spread out, shifted towards lower masses and lowered, leading to potentially large effects (\cf~Ref.~\cite{Kuhnel:2015vtw}).
\vs{3mm}

\noindent
{\it Bounds for Extended Mass Functions\;---\;}The derivation of bounds for extended mass functions is strictly speaking always subject to a specific model, such as the axion-curvaton model \cite{Lyth:2001nq, Kasuya:2009up, Kohri:2012yw}, the hybrid-inflation model \cite{Linde:1993cn, Clesse:2015wea}, or the running-mass model \cite{Stewart:1996ey, Leach:2000ea, Drees:2011hb, Drees:2011yz, Bugaev:2008gw, PhysRevD.83.083521}, just to mention a few (\cf~Ref.~\cite{Carr:2016drx} for an extensive overview). Therefore, it seems hopeless to draw any model-independent conclusions in this regard. However, it is indeed well possible at an approximate level. As has been pointed out recently by Green \cite{Green:2016xgy}, a (quasi) log-normal PBH mass function with derivative
\begin{align}
	\frac{\d n}{\d M}
		\equiv
								N\.\exp{\!
									\left[
										-
										\frac{( \log{M / M_{f}})^{2}}{2\.\sigma_{f}^{2}}
									\right]}
								\, ,
								\label{eq:dfdM}
\end{align}
fits a very large class of inflationary PBH models reasonably well (\cf~Refs.~\cite{Dolgov:1992pu, Dolgov:2008wu, Blinnikov:2016bxu} for the derivation and use of log-normal mass functions).\footnote{In Ref.~\cite{Green:2016xgy} it is noted that the least-squares fits of Eq.~\eqref{eq:dfdM} to the axion-curvaton and running-mass inflation differential mass functions $\d n / \d M$ differ from the original ones less than $10\%$, for parameters such that the PBH abundance peaks at $\sim 20\.M_{\odot}$.} {Above, $N$ is a normalisation constant chosen such that the integral of $\d n / \d M$ over all masses is equal to one. As a function of $M$, the \rhs~of Eq.~\eqref{eq:dfdM} is peaked at $M_{f}$; its width is controlled by $\sigma_{f}$.} In practice, for each pair of parameters $( \sigma_{f},\,M_{f} )$, a given set of constraints needs to be evaluated on the whole mass range, which then yields a limit on the PBH content.

The observational constraints which are included in our analysis are all of those stated in the caption of Fig.~\ref{fig:Constraints}, adapted for the extended mass function Eq.~\eqref{eq:dfdM} which we use throughout the rest of our paper.
\vs{3mm}

{\it Neutron-Star Capture:} Primordial black holes captured by neutron stars (NS) generically lead to the rapid destruction of the NS. Hence, the observations of neutron stars pose constraints on the PBH abundance. For a mono-chromatic mass spectrum, it has been argued in Ref.~\cite{Capela:2013yf} on the basis of a sufficiently large neutron-star survival probability that the constraint can be phrased as
\begin{align}
	1
		&\geq 
								f\,t_\text{NS}\.F_{0}
								\; ,
								\label{eq:constraint}
\end{align}
with the PBH dark-matter fraction $f \equiv \rho_{\rm PBH} / \rho_{\rm DM}$. Here, $t_{\rm NS}$ is the age of the star. The capture rate $F_{0}$ inside of a globular cluster (with core dark-matter density $\rho_{\rm DM, GC}$) reads
\begin{align}
	F_{0}
		&=
								\sqrt{6\pi\.}\;
								\frac{ \rho^{}_{\rm DM, GC} }{ M }\.
								\frac{ 2\.G_{\rm N} M_{\rm NS}\.R_{\rm NS} }
								{ \bar{v}\.( 1 - 2\.G_{\rm N} M_{\rm NS} / R_{\rm NS} )}\;
								\times
								\displaybreak[1]
								\notag
								\\[1.5mm]
		&\phantom{=\;}
								\times\!
								\left[
									1
									-
									\exp\!
									\left(
										-
										\frac{ 3\.E_{\rm loss} }{ M\.\bar{v}^{2} }
									\right)
								\right]
								,
								\label{eq:F}
\end{align}
with $G_{\rm N}$ being Newton's constant, $\bar{v}$ is the dispersion of the assumed Maxwellian velocity distribution of the PBHs, $R_{\rm NS}$ is the radius of the neutron star, and $M_{\rm NS}$ its mass. Finally, the average energy loss $E_{\rm loss}$ is approximately given by
\begin{align}
	E_{\rm loss}
		&\simeq
								\frac{ 58.8\.G_{\rm N}^{2} M^{2} M_{\rm NS} }{ R_{\rm NS}^{2} }
								\; .
								\label{eq:Eloss}
\end{align}

In the case of an extended mass distribution, the bound given in Eq.~\eqref{eq:constraint} changes. Let $I_{\rm bin} \in \Nbb$ denote the number of bins $\{ [ M_{i},\.M_{i + 1} ],\,i \leq I_{\rm bin}\}$. Then, for a given distribution specified by some $\d n / \d M$, we have
\begin{align}
	1
		&\gtrsim
								\sum_{i = 1}^{I_{\rm bin}}\.t_{\rm NS}\.f\.F_{0}^{(i)}
								\int_{M_{i}}^{M_{i + 1}}\!\d M\;\frac{ \d n }{ \d M }
								\; ,
								\label{eq:constraint-NS-capture}
\end{align}
where $F_{0}^{(i)}$ is evaluated on a mass within each bin, and the bin number $I_{\rm bin}$ should be chosen such that, to the precision sought, it does not matter where exactly in each bin the quantity $F_{0}$ is evaluated. For the evaluations of the constraint we use the parameter values $M_{\rm NS} = 1.4\.M_{\odot}$, $R_{\rm NS} = 12\.{\rm km}$, $\rho_{\rm DM, GC} = 2 \times 10^{3}\.{\rm GeV}\.{\rm cm}^{-3}$, $t_{\rm DM} = 10^{10}\.{\rm yr}$, and $\bar{v} = 7\.{\rm km}\.{\rm s}^{-1}$, as given in Ref.~\cite{Capela:2013yf}.
\vs{3mm}

{\it Pulsar Timing:} Ref.~\cite{Schutz:2016khr} argues that the abundance of $1$--$1000\.M_{\odot}$ PBHs might be considerably constrained via the non-detection of a third-order Shapiro time delay as the PBHs move around the Galactic halo. They present results of a respective Monte-Carlo simulation leading to a forecast for the Square Kilometre Array (SKA) which approximately follows an $f \sim M^{1 / 3}$ scaling. More precisely, and adopted to an extended PBH spectrum, the respective constraint might be written in the form
\begin{align}
	1
		&\gtrsim
								\sum_{i = 1}^{I_{\rm bin}}\.
								\left(
									\frac{ \tilde{M}_{i} }{ 1\.M_{\odot} }
								\right)^{\!\! 1 / 3}\,
								f
								\int_{M_{i}}^{M_{i + 1}}\!\d M\;\frac{ \d n }{ \d M }
								\; ,
								\label{eq:constraint-NS-capture}
\end{align}
where $\tilde{M}_{i}$ is some mass within each bin. As before, $I_{\rm bin}$ should be chosen sufficiently large.
\vs{3mm}

{\it Accretion Effects:} As first analysed by Carr \cite{1981MNRAS.194..639C}, in the period after decoupling, PBH accretion and emission of radiation might have a strong effect on the thermal history of the Universe. Ricotti {\it et al.}~\cite{Ricotti:2007au} have analysed this possibility in detail. In particular, it has been discussed that one associated effect of PBH is that they increase the optical thickness $\tau \ra \tau + (\Delta \tau)_{\rm PBH}$ which leads to
\begin{align}
	1
		&\gtrsim
								\sum_{i = 1}^{I_{\rm bin}}\.
			 					17.4
								\left(
									\frac{ \tilde{M}_{i} }{ 1\.M_{\odot} }
								\right)^{\!2}
								\,
								f
								\int_{M_{i}}^{M_{i + 1}}\!\d M\;\frac{ \d n }{ \d M }\;
								\, ,
								\label{eq:constraint-accretion-capture}
\end{align}
which is valid in the mass range $[ 30\.M_{\odot},\.10^{3}\.M_{\odot} ]$. In Eq.~\eqref{eq:constraint-accretion-capture}, we utilise the latest Planck constraint $(\Delta \tau)_{\rm PBH} < 0.012\;(95\%\,{\rm C.L.})$ as in Ref.~\cite{Chen:2016pud}.
\vs{3mm}

\noindent
{\it Other Bounds:} For the bounds from extragalactic $\gamma$-rays from evaporation, white-dwarf explosions, and lensing, we use a similar method to the one in Ref.~\cite{Carr:2016drx}. This allows to approximately determine, from constraints calculated assuming a delta-function halo fraction, whether an extended mass function is allowed. This methods utilises binning of the relevant mass range. Specifically, a given constraint $f_{c}$ is first divided into locally monotonic pieces. In each of these pieces one starts with the bin, say $i$, where the constraint is smallest, and integrates $\d n / \d M$ in this bin in order to obtain the fraction $f_{i}$. Then one goes to the next bin, integrating over $[ M_{i},\.M_{i + 1} ]$ in order to obtain $f_{i + 1}$, and so on. In the original formulation of Ref.~\cite{Carr:2016drx}, each $f_{j}$ is then compared to the largest value of $f_{c}$ in this bin (where the constraint is weakest). Recently, Green \cite{Green:2016xgy} pointed out that there may be cases where this procedure underestimates the constraint.

In this work, in order to avoid such a possible issue, instead of using the largest value of $f_{c}$, we will use the smallest one in each bin (where the constraint is weakest). This may overestimate the constraint, but for the smooth mass functions given by Eq.~\eqref{eq:dfdM} and the constraint used, by making the bins small enough, the error can be made arbitrarily small, and in particular much smaller than the error of using Eq.~\eqref{eq:dfdM} instead of the actual mass function of a given model.\footnote{We explicitly checked that all results in this work are {\it practically identical} irrespective of where in the bin constraint is evaluated. This is due to the large bin size and the smoothness of the mass function \eqref{eq:dfdM}. Of course, for a non-smooth mass function, special care has to be taken, but this is not relevant for the present analysis.}
\vs{2mm}

\noindent
{\it Results}{\;---\;}Figure \ref{fig:Parameter-Space} shows our main results for the constraints on the amount of PBH dark matter in the $( \sigma_{f},\,M_{f} )$-plane using the distribution given by Eq.~\eqref{eq:dfdM}. Here, the value of $M_{f}$, at which this distribution peaks, lies between $10^{-16}\.M_{\odot}$ and $10^{3}\.M_{\odot}$, and its width $\sigma_{f} \in [0.2,\,2]$. The region enclosed by the red-dashed contour in the middle of the plot around $10^{-9}\.M_{\odot}$ indicates the possibility for a PBH dark-matter fraction of $100\%$. This is not possible outside this narrow region, and excludes too narrow mass function ($\sigma \lesssim 0.4$), which in particular applies to the monocromatic case. Hence almost all of the parameter space does not allow for $100\%$ PBH dark matter, given the validity of the very stringent bounds in the mass range under investigation. As mentioned in the Introduction, it should be stressed that several of these constraints are based on rather uncertain assumptions (such as those deriving from accretion, neutron-star capture, or ultra-faint dwarfs), and might be weakened significantly once a more elaborated treatment of those has been performed. However, even if all the mentioned constraints are taken at face value, in the middle of the $M_{f}$ axis, there remains a region, and therefore a class of models with a necessarily extended mass function, which still allows for PBHs to constitute all of the dark matter.\footnote{In the process of finalising this paper, in Ref.~\cite{Niikura:2017zjd} additional microlensing constraints, using high-cadence observation of M31 with the Subaru Hyper Suprime-Cam, have been proposed. While these results still need to be investigated carefully, if reliable, they would put the currently most stringent upper bounds on the PBH abundance in the mass range $10^{-14}$ -- $10^{-6}\,M_{\odot}$. These bounds have very recently been studied in Ref.~\cite{Inomata:2017okj} using a method similar to the one used in this work. However, their Fig.~2, merely indicates whether $100\%$ PBH dark matter is excluded or not in the mentioned limited low-mass range, whereas our Fig.~\ref{fig:Parameter-Space} shows results for the whole mass range $10^{-18}$ -- $10^{3}\,M_{\odot}$, and furthermore indicates the exact value of the allowed PBH dark-matter fraction.}\vs{2mm}

\begin{figure}[t]
	\vs{-3mm}
	\centering
	\includegraphics[scale=1,angle=0]{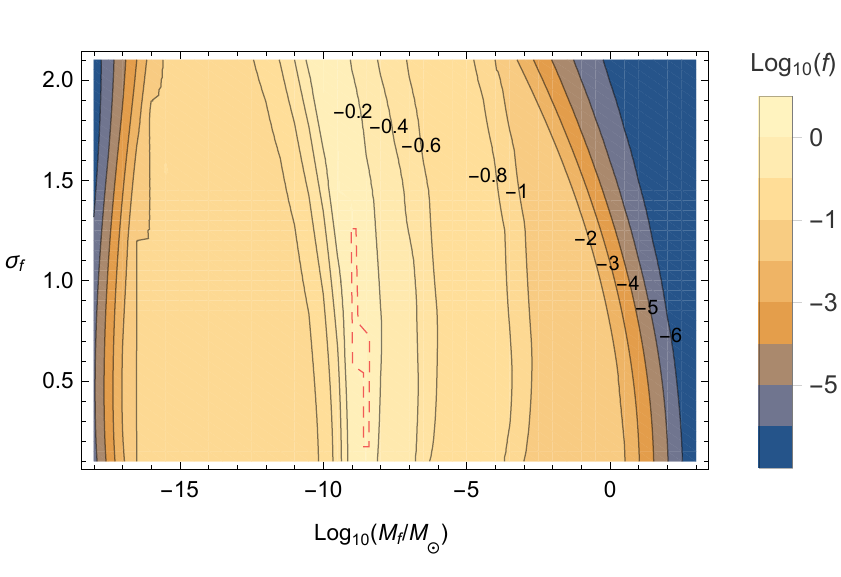}\;
	\caption{Maximum PBH dark-matter fraction $f \equiv \rho_{\rm PBH} / \rho_{\rm DM}$ as a function of 
		$\sigma_{f}$ and $M_{f}$ for the extended mass function specified in Eq.~\eqref{eq:dfdM}. 
		The region enclosed by the red-dashed contour in the middle of the plot indicates 
		the possibility of $100\%$ primordial black-hole dark-matter (see the legend). 
		The various contours are at $\log_{10}( f ) = 0\,\text{(red-dashed line)},\.
		- 0.2,\.- 0.4,\.- 0.6,\.- 0.8,\.- 1,\.- 2, \ldots,\.- 6$.\vs{-4mm}}
	\label{fig:Parameter-Space}
\end{figure}

\noindent
{\it Summary \& Outlook\;---\;}In this Letter we have presented the results of our systematic investigation of constraints for a wide class of extended mass function in the mass range $10^{-18}\.M_{\odot}$ to $10^{3}\.M_{\odot}$. {For these results, which are visualised in Fig.~\ref{fig:Parameter-Space}, very restrictive constraints on and forecasts for the allowed abundance of PBHs (as summarised in Fig.~\ref{fig:Constraints}) have been used. Hence, Fig.~\ref{fig:Parameter-Space} provides an approximate lower bound on the allowed PBH dark-matter fraction.}

{We confirm the results of Ref.~\cite{Carr:2016drx} that there still is a window in the mass range $10^{-10}\.M_{\odot}$ to $10^{-8}\.M_{\odot}$ which can accommodate for $100\%$ PBH dark matter.} Apart from the possibility of Planck-mass relics, to pose new constraints in the mentioned mass window seems to be crucial for providing an answer to the question whether primordial black holes can constitute the entirety of the dark matter.
\vs{10mm}

\acknowledgments
It is a pleasure to thank Alexander Dolgov, Benjamin Horowitz, Alexander Kashlinsky, and Peter Klimai for helpful comments. K.F.~acknowledges support from DoE grant DE-SC0007859 at the University of Michigan as well as support from the Michigan Center for Theoretical Physics. K.F.~and F.K.~acknowledge support by the Vetenskapsr{\aa}det (Swedish Research Council) through contract No.~638-2013-8993 and the Oskar Klein Centre for Cosmoparticle Physics.\\


\end{document}